\documentclass[12pt,a4paper,final]{iopart}

\usepackage{color}
\usepackage{iopams}  
\usepackage{graphicx}
\usepackage[breaklinks=true,colorlinks=true,linkcolor=blue,urlcolor=blue,citecolor=blue]{hyperref}
\usepackage{graphicx}
\usepackage[utf8x]{inputenc}
\usepackage[british]{babel}
\usepackage[margin=2cm]{geometry}
\usepackage{fancyhdr}
\usepackage{lastpage}
\usepackage{color}
\usepackage{eurosym}
\usepackage{rotating}
\usepackage{graphics}
\usepackage{cite}
\begin{document}
\title[{Interferom. length metrology for the dim. contr. of ultra-stable Ring Laser Gyroscopes}]{Interferometric length metrology for the dimensional control of ultra-stable Ring Laser Gyroscopes}
\author{J.~Belfi$^{1}$, N.~Beverini$^{1,2}$, D.~Cuccato$^{3,4}$, A.~Di Virgilio$^1$, E.~Maccioni$^{1,2}$,\\ A.~Ortolan$^{5}$, R.~Santagata$^{1,6}$}
\address{$^1$INFN Sezione di Pisa, Largo Bruno Pontecorvo 3, Pisa, Italy}
\address{$^2$Department of Physics, University of Pisa, Largo Bruno Pontecorvo 3, Pisa, Italy}
\address{$^3$Department of Information Engineering, University of Padova, Via Gradenigo 6/B, Padova, Italy}
\address{$^4$INFN Sezione di Padova, Via Marzolo 8, 35131, Padova, Italy}
\address{$^5$INFN National Laboratories of Legnaro, Viale dell'Universit\`a 2, Legnaro, Padova, Italy}
\address{$^6$Department of Physics, University of Siena, Via Roma 56, Siena, Italy}
\ead{\mailto{belfi@pi.infn.it}}

\begin{abstract}
We present the experimental test of a method for controlling the absolute length of the diagonals of square ring laser gyroscopes. The purpose 
is to actively stabilize the ring cavity geometry and to enhance the rotation sensor stability in order to reach the requirements for the
 detection of the relativistic Lense-Thirring effect with a ground-based array of optical gyroscopes. The test apparatus consists of two optical cavities 1.32 m in length, reproducing the features 
of the ring cavity diagonal resonators of large frame He-Ne ring laser gyroscopes. The proposed measurement technique is based on the use of a single diode laser, injection locked to a frequency stabilized He-Ne/Iodine frequency standard, and a single electro-optic modulator. The laser is modulated with a combination of three frequencies allowing to lock the two cavities to the same resonance frequency and, at the same time, to determine the cavity Free Spectral Range (FSR). 
We obtain a stable lock of the two cavities to the same optical frequency reference, providing a length stabilization at the level of 1 part in $ 10^{11} $, and the determination of the two FSRs with a 
relative precision of $\sim2\cdot10^{-7}$. This is equivalent to an error of $500\,\,\, \rm{nm}$ on the absolute length difference between the two cavities.
\end{abstract}

\section{Introduction}
Large frame optical gyroscopes based on He-Ne ring laser technology
are presently the most sensitive devices for measuring inertial rotational
motion\cite{Ullireview,stedreview}. The use of these systems has been considered in the past for several fundamental physics applications including 
experimental tests of fundamental symmetries \cite{tviolation}, axions detection \cite{axions}, test of metric theories of gravitation \cite{SCULLY, stedman2003}, 
and gravitational waves detectors \cite{GR_APPL}.  Recent improvements in the stability and resolution of this kind of devices fostered the GINGER 
(Gyroscope IN GEneral Relativity) project \cite{PRD_BOSI}. This experiment aims to develop a ground based array of large-frame laser gyroscopes (of at least 
$ 6\,\,\, \rm {m}$ in side) for the detection of the space-time geometry deformation induced by the Earth's rotation, 
also known as frame-dragging or Lense-Thirring effect.\\
The frequency difference $f_S$ between the two opposite traveling beams in a laser gyroscope is given by the basic theory of the Sagnac effect:
\begin{equation}
f_{S}=\frac{4\vec{A}\cdot \vec{\Omega}}{\lambda P}
\label {fSagnac}
\end{equation}
where $\vec{A}$ is the ring cavity area vector enclosed by the ring laser optical path, $P$ the cavity perimeter length, $\lambda$ 
the laser wavelength and $\vec{\Omega}$ the frame angular velocity. Errors and instabilities in the geometrical scale factor $ k_{S}=4A/\lambda P $ 
affect directly the measured rotation rate. The present limit of the most stable
devices is given by the slow deformations of the optical cavity
induced by temperature, pressure and aging of materials. Several approaches
have been followed in the last years in order to enhance the stability of the ring cavity geometry.

The best results have been obtained up to now by using monolithic
structures operating in very well isolated, i.e. passively stabilized,
laboratories. Today's best performing ring laser is the 4 m side G installed in a controlled area at the 
Geodetic Observatory of Wettzell, Germany  \cite{controllodiG3}. It is a monolithic block
of Zerodur with four mirrors mounted in optical contact with the cavity frame. It operated 
without any active control on the geometry until 2010. Since then, an active
control of the cavity shape has been engaged. Such a control system stabilizes the ring laser optical frequency against an optical frequency reference 
by tuning the pressure inside a vacuum tank enclosing the whole monolithic cavity \cite{controllodiG1,controllodiG2}. In this way
the cavity deformations induced by the variation of the pressure load
on the four optically contacted mirrors are canceled to first
order, with a consequent sensitive improvement of the long-term instrument stability. 

\begin{figure}[h!!!]
 \begin{minipage}[b]{8.5cm}
\includegraphics[width=7.0cm]{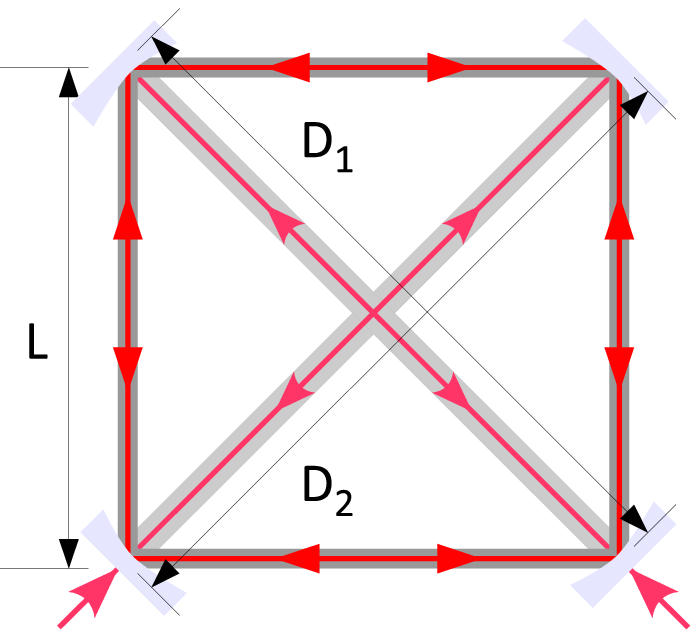}
\end{minipage}
\begin{minipage}[b]{8.5cm}
\includegraphics[width=8.5cm]{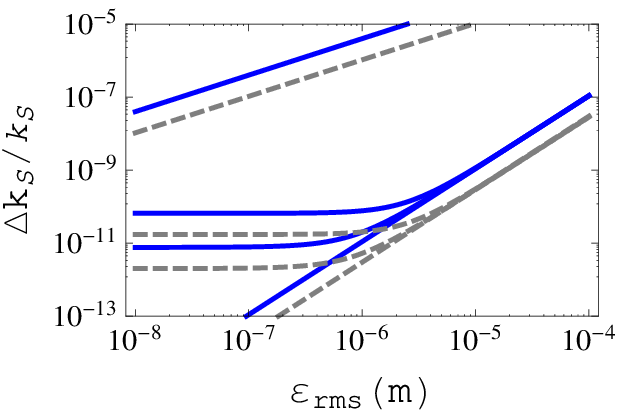}
 \end{minipage}
\caption{Left: Square ring laser gyroscope; diagonal Fabry-P\'erot cavities, probed by an external laser source, are shown. Right: Relative scale factor variations 
induced in a square ring laser by a displacement noise with root mean square amplitude $\varepsilon_{rms}$ acting on the four mirrors positions. 
Two cases with different side length $L$ and mirrors curvature radius $R$ are simulated: GP2 cavity, $ L=1.60\,\,\, \rm {m} $, $ R=4 \,\,\,\rm {m}$ (continuous blue line); GINGER cavity, $ L=6 \,\,\,\rm {m} $, $ R=6 \,\,\,\rm {m} $ (gray dashed line). For each cavity, the different plotted curves correspond to the following cases (from top to bottom): free geometry; diagonal lengths stabilized but unbalanced of $ 10^{-6}\,\,\, \rm {m} $; diagonal lengths stabilized but unbalanced of $ 10^{-7} \,\,\,\rm {m} $; equal absolute diagonal lengths. }
\label{stability}
\end{figure}

For a ring laser in a ground-based laboratory, the Lense-Thirring term is of order of 1 part in $10^{9}$ of the purely 
kinematic Sagnac term of equation~\ref{fSagnac}. The direct measurement of this effect implies the stabilization of the scale factor with an accuracy of at 
least 1 part in $10^{10}$. The stability of G is about one order of magnitude far from this target. 
A direct way to overcome the present limits is to increase the sensitivity of the ring by enlarging the cavity dimensions. However, it is not 
feasible to have a larger monolithic block of Zerodur; an alternative solution is to develop an active control acting on the corner mirrors positions. 
Experimental studies in this direction have been made on the G-Pisa ring laser \cite{G-Pisa}.
It has been shown that locking the laser emission frequency to a suitable optical frequency standard provides a well
defined configuration of the laser parameters but it is not the ultimate
solution to the problem of the cavity deformation. In fact, variations of the area
vector remain uncontrolled and the residual relative variation of the cavity side-lengths, even at constant perimeter,
introduce instabilities on the rotational signal by changing the backscattering phase. This last error source dominates 
the dynamics of small ring lasers (side lengths of about $1 \,\,\, \rm {m} $) and it is 
strongly reduced by increasing the gyroscope linear dimensions. For side length larger than $6\,\,\, \rm {m} $, the geometrical stability sets the main limit to the
system performances and can be controlled only by means of interferometric methods. 
The ideal instrument would be a cavity stabilized to the shape of a perfect square, where the four sides are perpendicular and of the same absolute length.
with a root mean square amplitude $\varepsilon_{rms}$, 
Both G and G-Pisa are square ring laser gyroscopes. One advantage of this cavity shape on the triangular
one is that the two diagonals can constitute additional degrees of freedom to constrain.
As a fact, typical He-Ne ring laser mirrors having a reflectivity of $99.999\%$ at $45 \,\,\, \rm {deg}$ angle of incidence, 
still have a good reflectivity at normal incidence. This allows us to use the same external laser source providing the perimeter reference frequency
for stabilizing the length of the two Fabry-P\'erot cavities formed by the opposite mirrors of the ring. By performing a beam steering
simulation of the ring cavity it can be shown \cite{GP2_1} that, when the length
of the two diagonals is locked to the same value, the perturbations to the mirror positions affect only quadratically
the ring laser scale factor. In figure~\ref{stability} it is shown the simulated relative variation of the scale factor$k_{S}=4A/(\lambda P)$ 
induced by a random displacement noise applied to the four mirrors centers of curvature. The position of each mirror has been changed 
along the three directions of space by three independent random processes with root-mean-square amplitude $\varepsilon_{rms}$; 
the case of a cavity with equal absolute diagonal lengths is compared to the uncontrolled case and to the case where 
the two lengths are both stabilized better than 1 part on $10^{11}$ but differ one from the other by a quantity $\delta D$. Details are in the caption.

It is worth noting that even in the case of unbalances of the order of $1 \,\,\, \rm{\mu m}$ the length 
stabilization constraint provides a high rejection of the 
mirrors position perturbations from the scale factor, with residual fluctuations at level of 1 part in $10^{10}$. 
Motivated by this numerical results we designed a new ring laser gyroscope, called GP2 \cite{GP2_2}, 
dedicated to the optimization and stabilization of 
the geometrical form factor and, in particular, to the experimental implementation 
of the length stabilization of the diagonal cavities. It is a square resonator $1.6 \,\,\, \rm {m}$ in side, equipped with a system of 
6 piezoelectric actuators controlling the four mirrors positions. A special vacuum chamber encloses the four mirrors, the optical path along the ring cavity as well 
as the two diagonals.\\

In this paper we present an experimental technique to set equal the
lengths of two linear Fabry-P\'erot resonators formed by curved mirrors used typically as corner reflectors
in He-Ne ring laser gyros, in order to simulate the ring cavity diagonals on an optical bench. The measurement of the cavity absolute
length can be obtained with very high precision ($<10^{-11}$) if the cavity resonance is locked to the frequency of an optical frequency standard and 
the integer number $n$ of half-wavelengths contained between the two mirrors is unambiguously estimated. 
This last can be obtained by measuring the cavity FSR with a precision 
better than $1/n$. In the recent
literature, particularly in the field of gravitational wave detectors,
several techniques have been proposed to measure the FSR of long linear
cavities. In particular, Araya et al.\cite{Araya} measured the absolute
length of an approximately $300\,\,\rm{m}$ long Fabry-P\'erot cavity by FSR measurement, using two EOM for frequency modulation and attaining a measurement uncertainty
of $10^{-8}$; Aketagawa et al. \cite{Aketagawa} measured the FSR
of an approximately 0.170 m long cavity using a single EOM with an uncertainty of $10^{-6}$. For a more detailed list of developed techniques see also \cite{Aketagawa2}. 

The strategy we adopted is based on the use of a single frequency-stabilized diode laser at $633\,\,\rm {nm}$, phase modulated
with a single EOM driven by a combination of three independent modulation frequencies. The first modulation provides the Pound-Drever-Hall signals for locking the two cavities to the same laser optical frequency. 
The second modulation creates a set of sidebands spaced by a harmonic of the FSR frequency. The third modulation is a small dithering of the second frequency 
for shifting the FSR resonance detection down to few tens kHz. 

The optical reference frequency in the experiment is provided
by a He-Ne laser, frequency stabilized on the saturated absorption
line R-127 11-5 of Iodine. The accurate determination of the FSR is measured by locking a VCO (Voltage Controlled Oscillator) 
to the center of the cavity dynamic resonance and by counting the frequency with a counter whose stability is of $ 1\,\, \rm {ppb} $.\\
In section 2 we give a theoretical description of the measurement technique. In section 3 we present the experimental apparatus including the 
laser source at $ 633\,\, \rm {nm}$. Section 4 reports the discussion of the results concerning the stabilization of the cavity length, the 
estimate of the free-spectral-ranges of the two cavities and their absolute length difference.

\section{Measurement principle}
\label{theory}
Let's consider a linear cavity formed by
two concave spherical mirrors of radius $R$ separated by a distance $L$. The frequency $f_{n}$ of a $TEM_{00}$ laser beam resonating in the longitudinal mode of order
$n$ is given by:

\begin{equation}
\label{EQ1}
f_{n}=\left(\frac{v}{2L}\right)\left[n+\frac{1}{2\pi}(\Psi_{R}+\Phi_{n})\right]
\end{equation}
where $v$ is the speed of light in the dielectric medium contained in the cavity, $\Psi_{R}=2\cos^{-1}\left[1-(L/R)\right]$ the Gouy-phase correction, and
$\Phi_{n}$ a frequency-dependent phase shift correction due to
the dielectric coating of the mirrors. The cavity length $L$ can be measured with an accuracy ultimately limited by the knowledge of the reference laser frequency 
if the cavity is locked to the laser and the integer number $n$ is estimated together with $\Psi_{R}$ and $\Phi_{n}$. 
$\Psi_R$ can be obtained with high accuracy from the knowledge of $R$,
whereas the dielectric shift contribution term, depending on the mirror manufacturing properties, can be evaluated from the reflectivity 
curve of the mirrors \cite{Hood}. However, for two cavities having the same kind of mirrors (i.e. produced in the same run of coating process) 
we can assume the two $\Phi_{n}$ to be equal.

For each cavity  we require two error signals in order to lock the cavity resonance to the laser carrier frequency and, subsequently, to lock the phase modulation
frequency to a harmonic of the FSR. The laser electric field injected into the two cavities has the form:
\begin{eqnarray}
E_{in}(t) & = & E_{0}e^{i\left\{ \omega_{0}t+\alpha\sin\omega_{A}(t)+\beta\sin\left[\left(\omega_{B}+\triangle\sin\omega_{C}t\right)t\right]\right\} }\\\nonumber
& = & E_{0}e^{i\omega_{0}t}\left[\left(-1\right)^{p}\sum_{p}J_{p}\left(\alpha\right)e^{i p\omega_{A}t}\right]\left[\left(-1\right)^{q}
\sum_{q}J_{q}\left(\beta\right)e^{iq\left(\omega_{B}+\triangle\sin\omega_{C}t\right)t}\right]
\label{eq_field}
\end{eqnarray}
here
\begin{itemize}
\item $ \omega_{0} $ is the optical frequency;
\item  $\Delta\omega_{cav}<\omega_{A}<FSR$ is the Pound Drever Hall modulation frequency for the carrier lock, where $\Delta\omega_{cav}$ is the cavity linewidth;
\item $\omega_{B}=m FSR$ is the modulation frequency for the cavity dynamic resonance excitation; 
note that the error on the resonance frequency measurement nominally decreases with the integer $m$  as  $m^{-1} $;
\item  $\omega_{c}\sim 30\,\,\, \rm{kHz}$ is the frequency of the dithering applied to $\omega_{B}$ for the lock-in detection of the
FSR resonance signal.\\
\end{itemize}

$J_{p}(\alpha)$,  $J_{q}(\beta)$ are first kind Bessel functions, $\alpha$ and $\beta$ are the modulation indexes 
for the modulations at $\omega_{A}$ and $\omega_{B}$, and $ \Delta $
is the frequency deviation of the dithering at $\omega_{C}$. For
small values of $\alpha$ and $\beta$, such that almost all the power is in the carrier and the first order sidebands, only the
first sidebands of the modulated spectrum (p=-1,0,1 and q=-1,0,1) can be considered.
Thus the incident field can be written in terms of 9 modes,
i.e. the carrier optical frequency and two group of four sidebands, as
shown in figure~\ref{spectrum}. 
\begin{figure}[h!!!]
\label{spectrum}
\centering
\includegraphics[width=12cm]{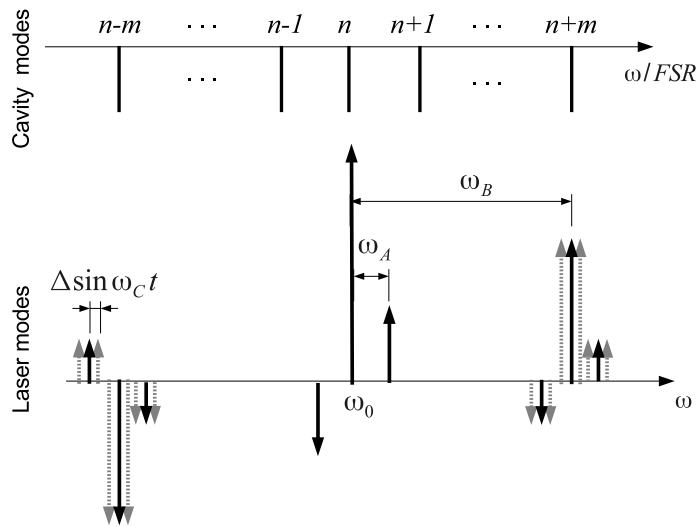}
\caption{Schematic diagram of the cavity modes and the laser spectral structure. The frequency dithering at $\omega_C$ is not applied to the central pattern of 
three modes, but only to the six extreme modes.}
\label{spectrum}
\end{figure}
Within this representation we can write:
\begin{equation*}
E_{in}(t)=E_{0}\sum_{k=1}^{9}A_{k}e^{i\omega_{k}t},
\end{equation*}
where the 9 pairs of $A_{k}$, $\omega_{k}$ are defined in table~\ref {components} of Appendix. The reflected field $E_{r}(t)$
is obtained by applying the cavity reflection transfer function $R\left(\omega_{k}\right)$, defined in Appendix (equation~\ref {eq:Rtransf}), 
to each of the $ k_{th} $ field components:

\begin{equation*}
E_{r}(t)=E_{0}\sum_{k=1}^{9}A_{k}R\left(\omega_{k}\right)e^{i\omega_{k}t}.\label{eq:reflected beam}
\end{equation*}
The signal detected by the photodetector is proportional to
the reflected field intensity $I_{r}\propto|E_{r}|{{}^2}$, i.e.:

\begin{equation*}
I_{r}(t) \propto E_{0}^{2}\sum_{k,k'=1}^{9}A_{k}A_{k'}R\left(\omega_{k}\right)R^{*}\left(\omega_{k'}\right)e^{i\left(\omega_{k}-\omega_{k'}\right)t},\label{eq:refl.intensity}
\end{equation*}
where $ R^{*} \left(\omega_{k}\right) $ is the conjugate of the cavity reflection transfer function. 

If we select, by means of a phase sensitive detection technique, the
components of the signal oscillating in phase with $\sin\omega_{A}t$
and $\sin\omega_{C}t$, see~\ref{omegaA} and~\ref{omegaC}, we obtain the
error signal for the carrier $\epsilon_{0}(\omega)$ and for the sidebands
$\epsilon_{s}(\omega)$: 
\begin{eqnarray*}
\epsilon_{0}(\omega) & \propto & \sum_{i=1,4,7}2A_{i}A_{i+1} Im
[S_{i+1}]\\
\epsilon_{s}(\omega) & \propto & 
\sum_{i=1,2,3,7,8,9}A_{i}^{2}\frac{d|R(\omega)|^{2}}{d\omega}\Big|_{(\omega=\omega_{i},\Delta=0)}
\label{eq:Err.Signals}
\end{eqnarray*}

where
\begin{equation}
S_{i}=R(\omega_{i-1})R^{*}(\omega_{i})-R(\omega_{i})R^{*}(\omega_{i+1}).
\label{eq:Si}
\end{equation}

\section{Experimental apparatus}
The block scheme of the diagonal cavities test bench is shown in figure~\ref{detection}. In the following we describe the different parts of the apparatus:
laser source, injection optics, test cavities and readout electronics.

\begin{figure}[h!!!]
\label{detection}
\centering
\includegraphics[width=15cm]{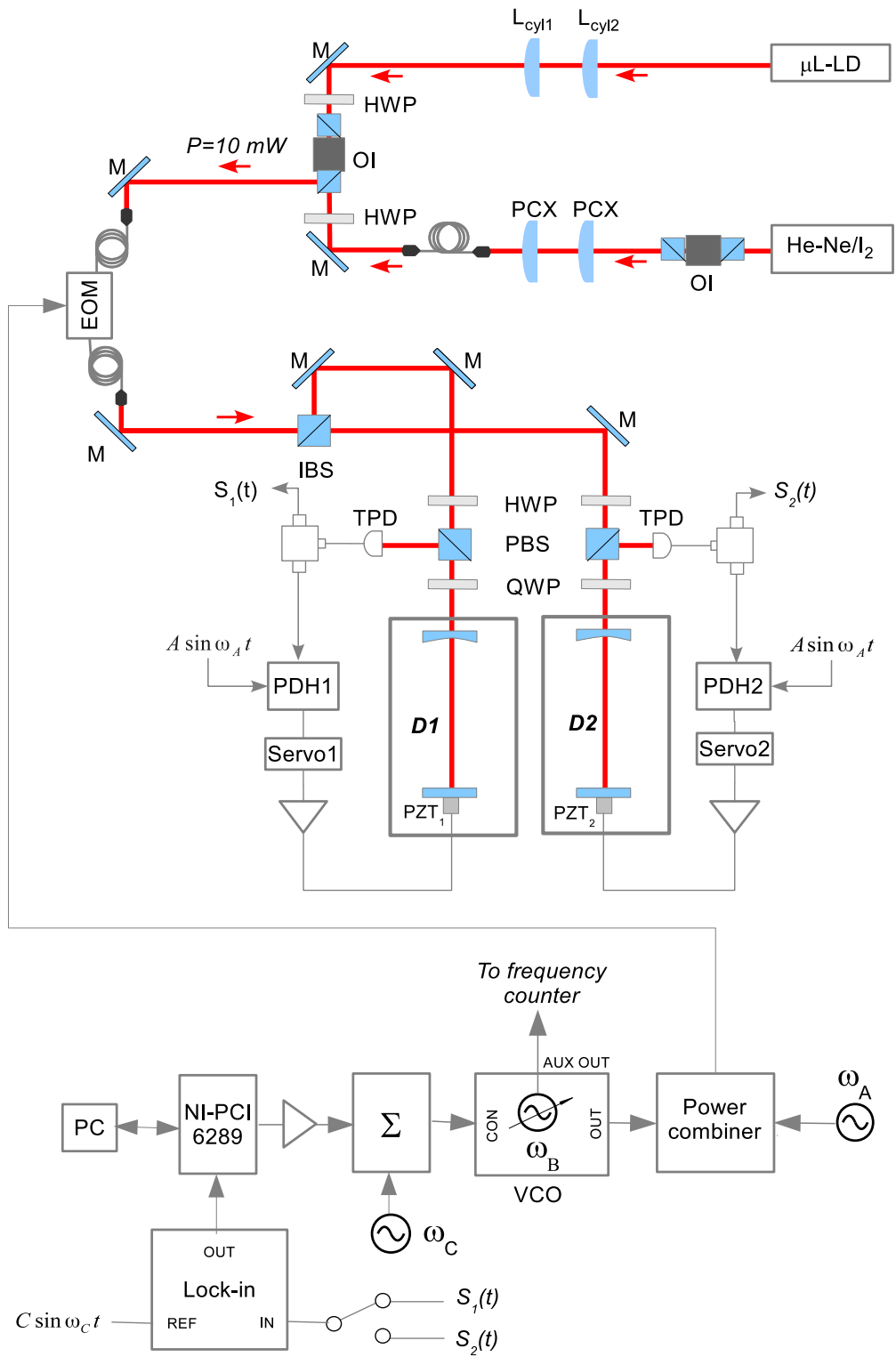}
\caption{Block scheme of the experimental apparatus. The upper part represents the optical breadboard with the injection locking setup. $\rm{\mu L-LD}$: micro-Lens Laser Diode,
 M: Mirror, HWP: Half Wave Plate, QWP: Quarter Wave Plate, PBS: Polarizing Beam Splitter, IBS: Intensity Beam Splitter, OI: Optical Isolator, VCO: Voltage Controlled Oscillator, TPD: Transimpedance Photodiode, PDH: Pound Drever Hall detection stage.}
 \label{detection}
\end{figure}

\subsection{Laser source}
In order to achieve the desired long term stability of the geometrical parameters of the gyroscope, a stable laser source with good spectral purity is needed. 
The reference frequency is provided by an iodine stabilized  He-Ne laser.
Typical output power of this systems is of the order of $100 \,\,\rm{\mu W}$. Since for our application a higher power is required, following the approach of \cite{VANIO}, 
we implemented a light amplifier based on injection-locking, so that the spectral purity of the reference 
laser is transfered to a $10\,\,\rm{mW}$ diode laser. 

\subsubsection{Reference laser}
The iodine-stabilized He-Ne laser is a master-slave system itself and it was provided by the Institute of Laser Physics of 
Novosibirsk, Russia. Here a He-Ne laser (slave) is phase-locked, with an offset frequency of $ 8 \,\,\rm {MHz} $, to another
He-Ne laser (master). The frequency of the master laser is locked to one of 14 vibro-rotational optical transitions of molecular $^{127}I_{2}$ contained in a fused
silica intracavity cell. The control system is based on the third derivative detection of the
saturated absorption signal. 

The principle of the stabilization is to keep the laser wavelength resonant to the chosen transition by adjusting the length of the cavity.
To this purpose the master laser has two piezoelectric transducers acting on the position of the two cavity mirrors. A fast PZT is used to compensate 
the fast fluctuations and apply a $5 \,\,\rm {kHz}$ modulation for the third derivative detection and a slow PZT compensates the thermal and mechanical frequency drifts. 
The relative Allan deviation of this laser reaches the level of $10^{-11}$ at $100 \,\,\rm {s}$.

\subsubsection{Probe laser}
The light source, acting as slave oscillator in the injection locking
setup, is a diode laser emitting at 633 nm. Commercially available chips at this wavelength  
are quite unreliable since they are at the limit of the wavelength range of red diode lasers. As a consequence, in 
most of the cases their performances degrade very rapidly in time and typically require to be cooled down to several degrees below zero. 
Following \cite{VANIO} we selected the microlens-coupled diode laser chip (Blue Sky Research
PS010). This system operating at $4.5^{\circ}C$  provides a single-mode output power of about $10 \,\  \rm {mW} $ 
resonant with the He-Ne transition. The chip is mounted inside a cylindrical aluminium assembly. A Peltier element and a thermistor are used to stabilize the 
temperature junction within $1 \,\ \rm {mK} $. The integrated microlens provides a nearly circular and Gaussian distribution of the intensity profile, 
making it possible to easily achieve a good mode matching with the He-Ne master laser.

\subsection{Injection locking}
The light of the reference laser is fiber coupled and
injected into the slave laser through the output port of an optical
isolator, as described in \cite{VANIO}. The typical master laser 
power entering the isolator is $P_{inj}\sim 100  \,\ \rm {\mu W}$.
The scheme of the injection locking setup is presented in the upper part of figure~\ref{detection}.
To get an efficient coupling, the spatial mode of the injected beam must match the mode of the slave laser.
We make use of a couple of cylindrical lenses in order to mode match the two lasers. The tuning of the free-running diode laser frequency is obtained
by adjusting the junction temperature (coarse) and current (fine). At diode laser working point 
$(T,I_{diode})=(4.5  \,\ \rm {^{\circ}C}, 65  \,\ \rm {mA})$ we obtain a typical locking-range \cite{siegman} of $\sim 10 \,\ \rm {GHz} $. 

\subsection{Ring laser diagonal cavities test-bench}
A drawing of the mechanical assembly of the Fabry-P\'erot cavities used in our set-up is shown in figure~\ref{cavita}. 
Two cavities are mounted onto an optical table equipped with a pneumatic isolation system. 
The two resonators consist of two spherical mirrors mounted on tip-tilt holders rigidly connected by Invar spacers.
A piezoelectric translator is attached to the output mirror in order to scan the cavity and implement the feedback correction to the cavity length. 
The piezoelectric transducer is a hollow cylinder $20.3\,\,\rm{mm}$ in length and diameter, $1.5\,\,\rm {mm}$ in thickness. 
The measured mechanical response is $ 2.2   \,\ \rm {n m/V}$ (full range voltage: $ 700   \,\  \rm {V} $). 
While the output mirrors have equal nominal specifications, the reflectivity of the input mirrors are different, 
leading to different cavity quality factors. In figure~\ref{spettrofotometro} 
is reported the transmission of a ring laser supermirror when operated at normal incidence. The parameters of the two Fabry-P\'erot resonators 
are summarized in table~\ref{fp}. 

\begin{figure}[h!!!]
\centering
\includegraphics [width=14cm] {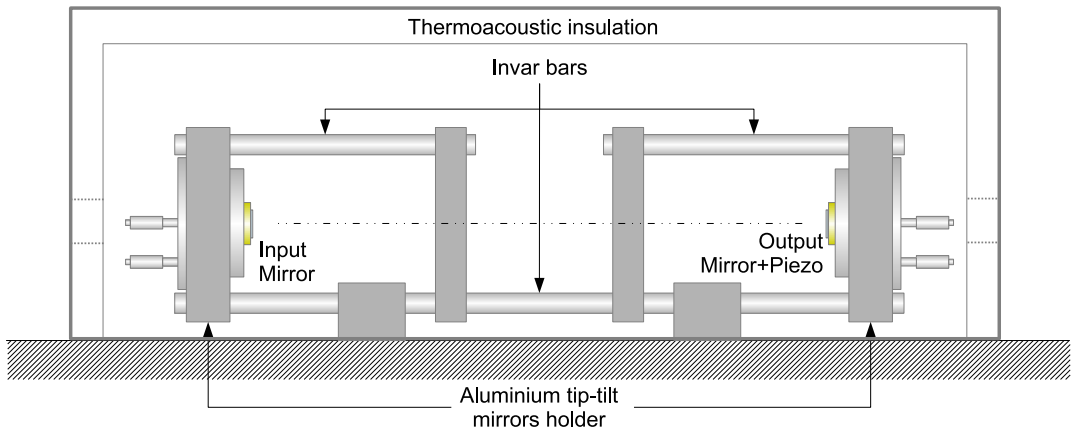}
\label{cavita}
\caption{Mechanical assembly of the Fabry-P\'erot cavities used in the test experiment.}
\label{cavita}
\end{figure}

\begin{table}[h]
\label{fp}
\centering
\begin{tabular}{|l|l|l|}
\cline{1-3}
Parameter & Cavity1 & Cavity2 \\ \cline{1-3}
$R$ & $ 4 \,\,\rm {m} $ & $ 4 \,\,\rm {m} $\\ \cline{1-3}
$D$ & $\simeq$$ 1.32\,\, \rm {m} $ &  $\simeq $$ 1.32\,\, \rm {m} $ \\ \cline{1-3}
$FSR$ & $\simeq$ $ 113\,\, \rm {MHz} $ & $\simeq$ $ 113\,\, \rm {MHz} $ \\ \cline{1-3}
$\Psi_{R}$ & $\simeq$$ 0.27 $ & $\simeq$ $ 0.27 $ \\ \cline{1-3}
$r_{in}^{2}$ & $ 0.988 $ & $ 0.997 $ \\ \cline{1-3}
$r_{out}^{2}$ & $ 0.999 $ & $ 0.999 $ \\ \cline{1-3}
$\mathcal{F}$ & $ 504 $ & $ 2094 $\\ \cline{1-3}
$FWHM$ & $ 225\,\, \rm {kHz} $ & $ 54 \,\,\rm {kHz} $ \\ \cline{1-3}
\end{tabular}
\caption{Optical parameters for the two resonators emulating the diagonals of a ring laser of $1.32 / \sqrt{2}\sim 0.93 \,\,\rm{m}$ in side length.}
\label{fp}
\end{table}

\begin{figure}[h!!!]
\label{spettrofotometro}
\centering
\includegraphics[width=14cm]{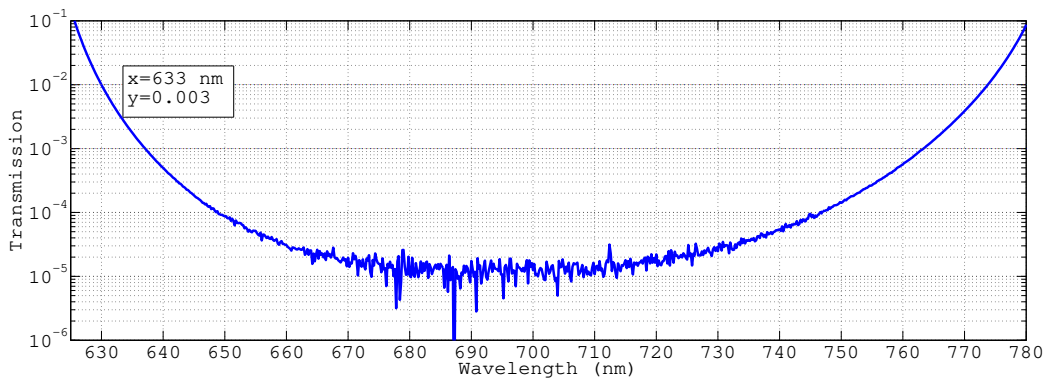}
\caption{Measured transmission spectrum for the input mirror of cavity 1. }
\label{spettrofotometro}
\end{figure}

\subsection{Readout electronics}
Before entering into the two cavities, the injection-locked diode laser beam is phase-modulated with a combination of 3 frequencies as reported in equation~\ref{eq_field},
 with $\omega_{A}=30\,\,\rm{MHz}$, $\omega_{B}\simeq680 \,\,\rm{MHz}$, $\omega_{C}=33 \,\,\rm{kHz}$. This is done by means of a single
fiber-coupled electro-optical phase modulator (EOM, Model PM635,
Jenoptik). The reflected light from each cavity is detected with a transimpedance photodiode. The photodiode output is split in two lines using a power divider.
One output of the power divider is demodulated at $\omega_{A}$ according to the standard Pound-Drever-Hall scheme 
in order to obtain an error signal $\epsilon_0(\omega)$ for the carrier lock. 
The second output of the divider is phase-detected by a digital lock-in amplifier ($EG\&G$ 7260) referred to the modulation frequency $\omega_C$.
The amplitude of the reflected signal in phase with the modulation is used as error signal 
$\epsilon_S(\omega)$ for the sideband lock. Simulated and measured error signals are shown in figure~\ref{ES_C} and figure~\ref{ES_S} respectively. 

\begin{figure}[h!!!]
 \centering
 \includegraphics[width=15cm]{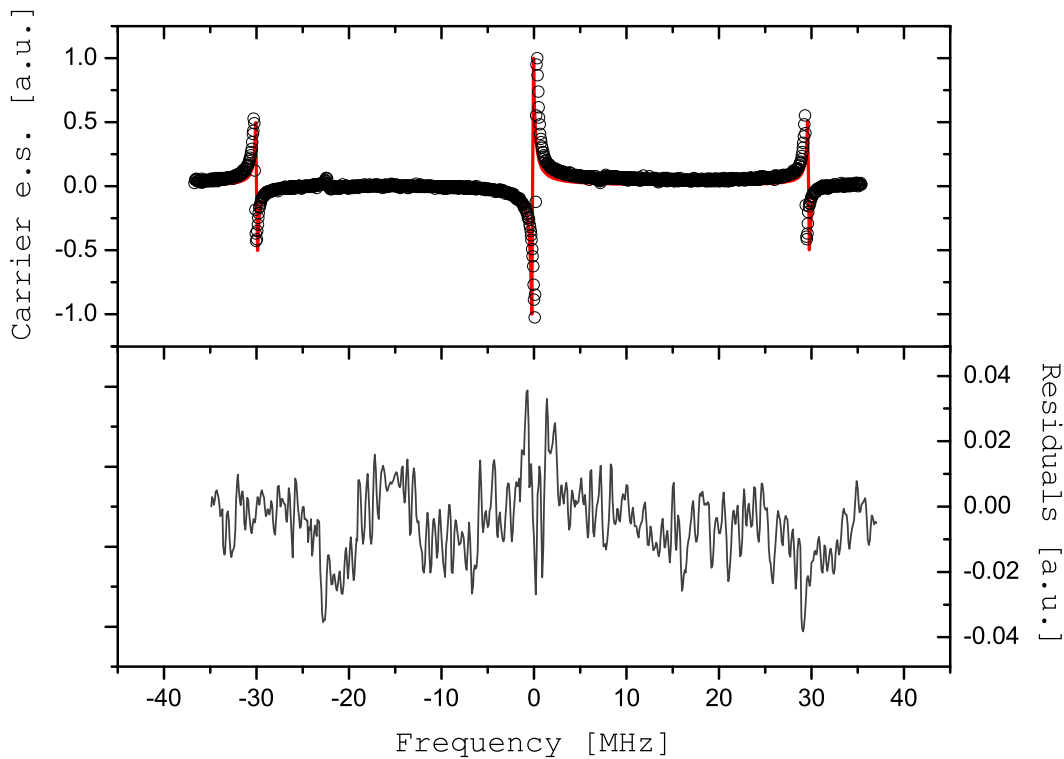}
 \label{ES_C}
\caption{Above: simulated (continuous red line) and measured (black circles) carrier error signal versus $\omega_0$, when $\omega_B \sim 6\cdot FSR$. Modulation parameters: 
$\alpha=0.7$, $\omega_A=30 \,\,\rm {MHz}$, $\beta=1$, $\omega_C= 33\,\, \rm{kHz}$, $\Delta\sim 200\,\, \rm{kHz}$. Sweep time: $ 100 \,\,\rm {ms}$. Below: residuals of the measured signal from the simulated one.}
\label{ES_C}
\end{figure}

\begin{figure}[h!!!]
 \centering
 \includegraphics[width=15cm]{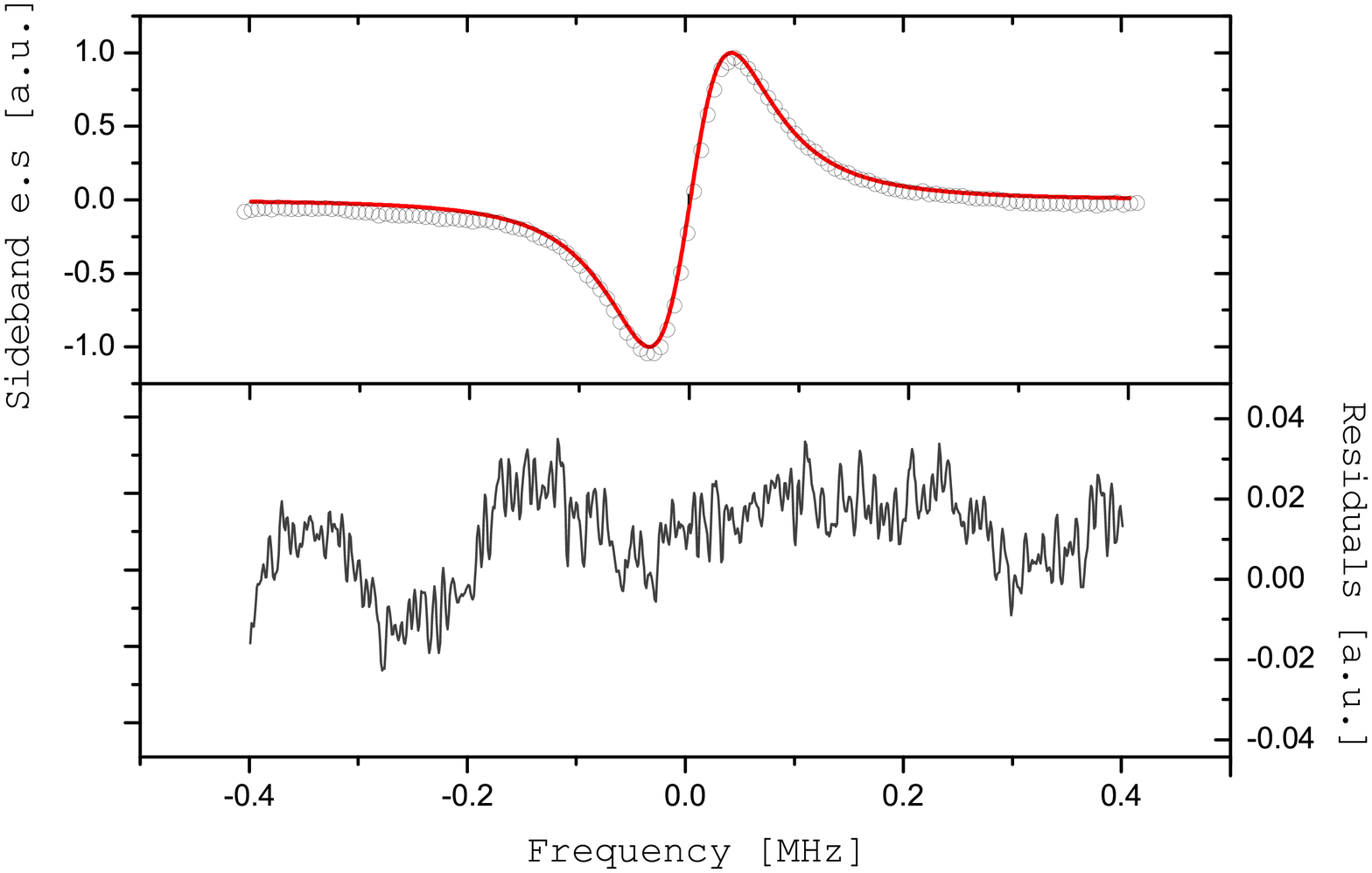}
 \label{ES_S}
\caption{Above: simulated (continuous red line) and measured (black circles) sideband error signal versus $\omega_B$. Modulation parameters: $ \alpha=0.7 $, $\omega_A= 30 \,\,\rm{MHz}$, $ \beta=1 $, $\omega_B= 680\,\, \rm{MHz}$, 
$\omega_C= 33\,\, \rm{kHz}$, $\Delta\sim 200\,\, \rm{kHz}$. 
Lock-in sensitivity: $ 10 \,\,\rm{V}/1 \,\,\rm{mV} $, lock-in time constant: $ 640 \,\,\rm{\mu s} $, sweep time: $ 5\,\, \rm {s} $. Below: residuals of the measured signal from the simulated one.}
\label{ES_S}
\end{figure}

\section{Results}
\label{results}
The two error signals $\epsilon_0$ and $\epsilon_S$ are used to apply corrections respectively to the cavity PZTs and to the VCO generating the sidebands at 
$\omega_B\sim 6\cdot FSR$. The feedback loops controlling the cavity mirrors displacement are two analog PI circuits with a bandwidth of about $ 1\,\, \rm {kHz} $.
The residual noise in the error signal expressed as equivalent displacement noise for the two stabilized cavities is shown in figure~\ref{PSD}. 

 \begin{figure}[h!!!]
 \centering
\includegraphics[width=16cm]{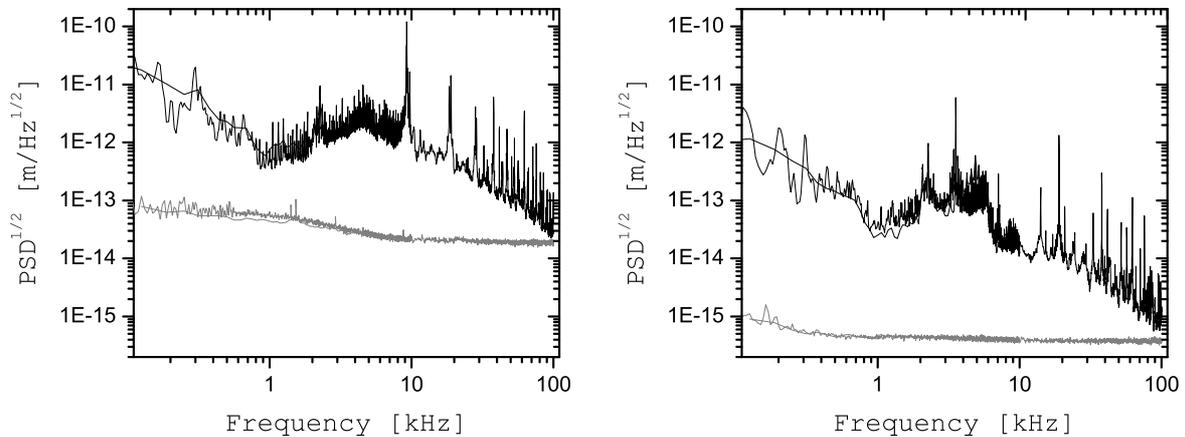}
 \label{PSD}
\caption{Power spectral density of the residual displacement noise for the cavity 1 (left) and cavity 2 (right). Gray lines are the electronic 
noise levels measured out of resonance.}
 \label{PSD}
 \end{figure}
\begin{figure}[h!!!]
 \begin{minipage}[b]{8.5cm}
\includegraphics[width=8.5cm]{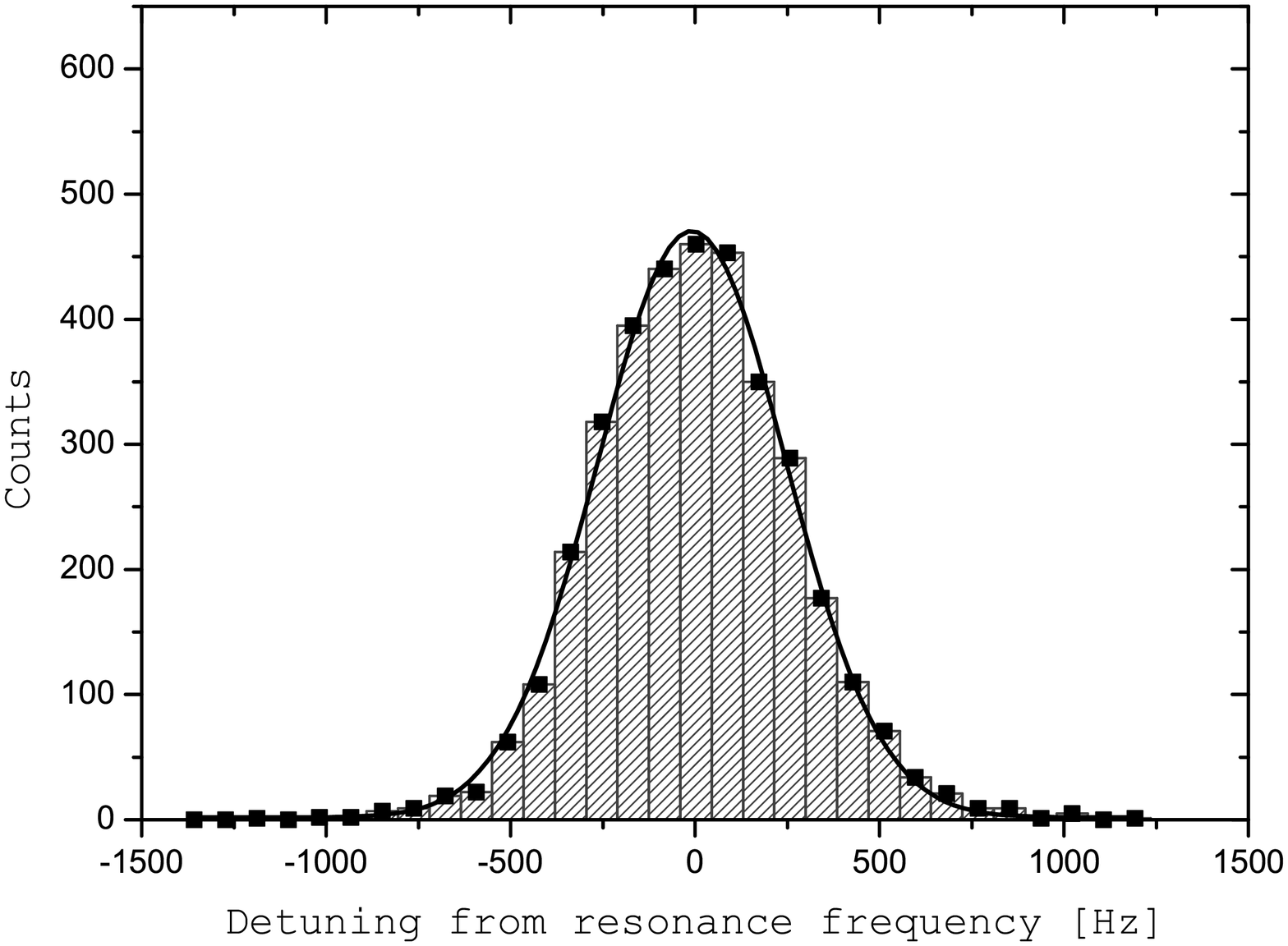}
\end{minipage}
\begin{minipage}[b]{8.5cm}
\includegraphics[width=8.5cm]{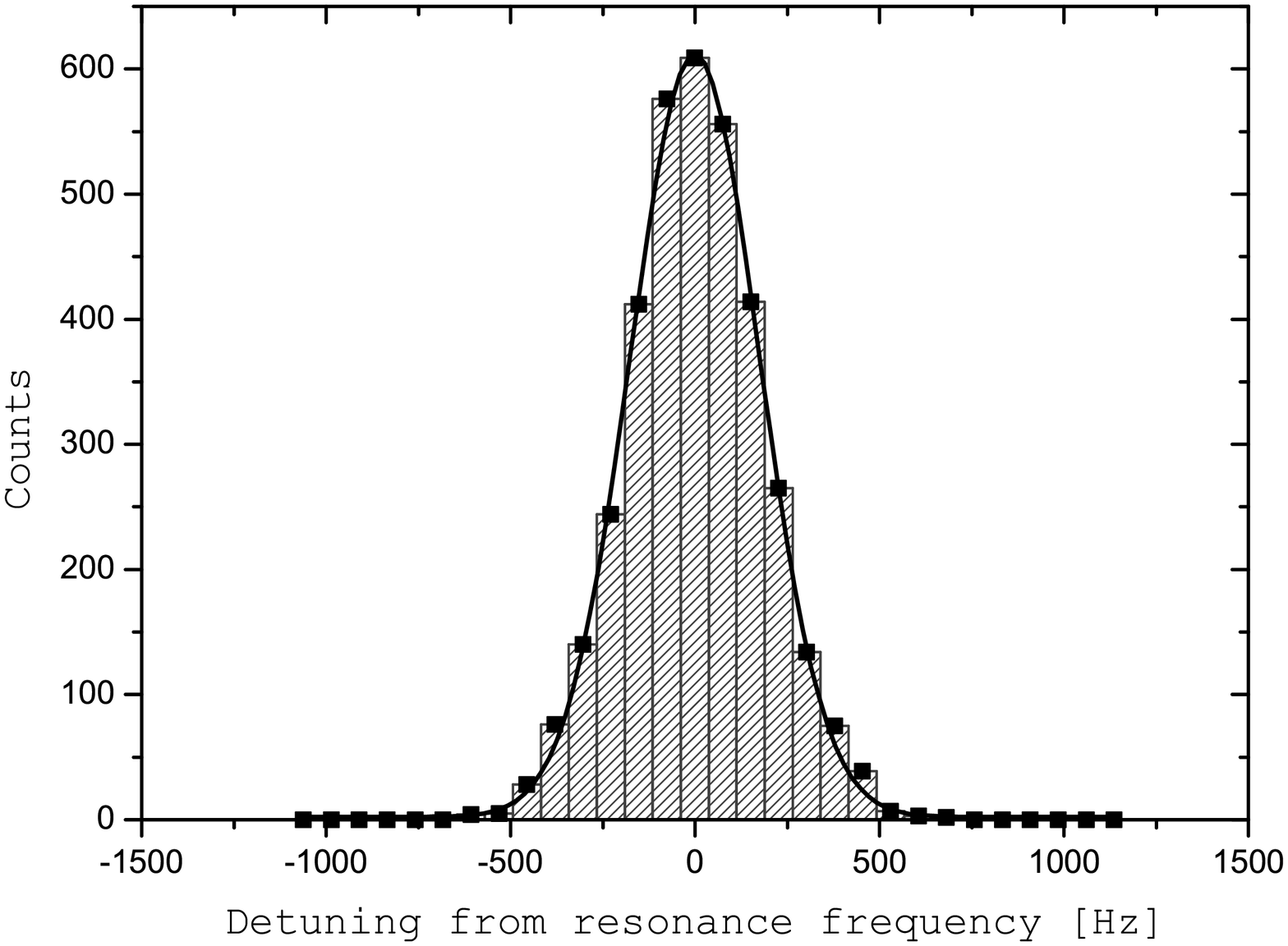}
 \end{minipage}
\caption{Frequency counting distribution for the measurements of $6\cdot FSR$. Total measurement time: $ 140 \,\ \rm {min} $. 
Frequency counter gate-time: $ 1\,\, \rm {s} $. From the gaussian fits on the two distributions we obtain an estimate for the central frequency $f_c$ and the standard deviation
$\sigma$. Cavity 1 (left): $f_{c1}=681211560\,\, \rm{Hz}$, $\sigma_1=230\,\, \rm{Hz}$, $\chi_R^2=0.997$. Cavity 2 (right): $f_{c2}=680000798\,\, \rm{Hz}, \sigma_2=140\,\,\rm{Hz}, \chi_R^2=0.998$}
\label{gaussiane}
\end{figure}

In figure~\ref{feedback} we report the acquisition of the feedback signals controlling 
the two cavity lengths. In the same graph is also shown the room temperature variation. A drift of the cavity mechanics is clearly visible in addition to 
the effect of thermal expansion.
\begin{figure}[h!!!]
 \centering
 \includegraphics[width=12cm]{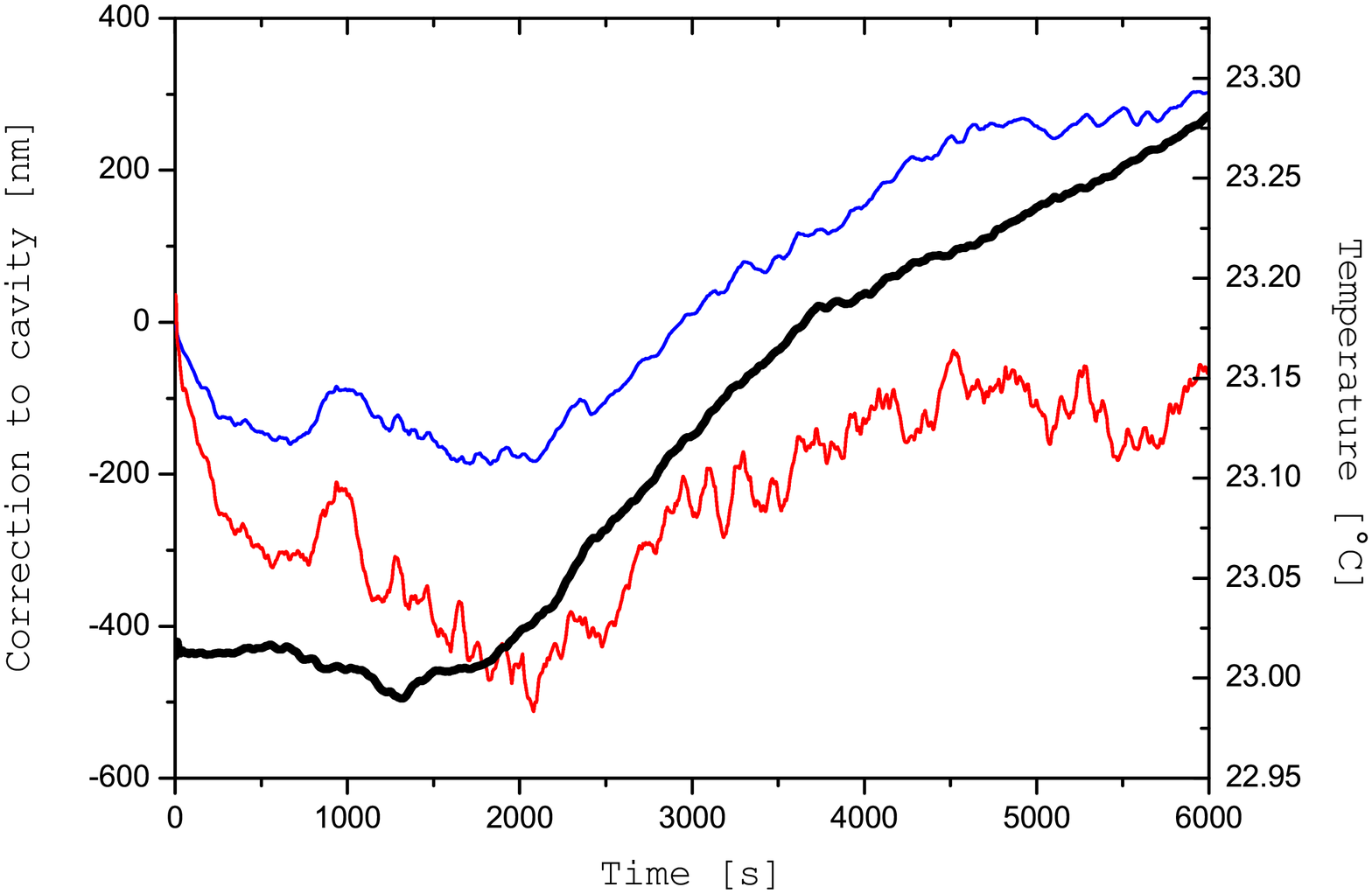}
 \label{feedback}
\caption{Thin traces: closed-loop corrections to the cavities output mirror positions (blue line: cavity 1; red line: cavity 2). Thick trace: room temperature.}
 \label{feedback}
 \end{figure}
Once the two cavity resonance frequencies are locked to the laser carrier frequency, the error signals $\epsilon_{S(1,2)}$ are processed by a LabView program running on 
a PC equipped with a DAQ board (NI-PCI 6289).
The system locks the VCO frequency alternatively to the 6-th harmonic of $FSR_1$ or $FSR_2$ at about $ 680 \,\,\rm{MHz} $. 
As one FSR is determined with the required resolution, the VCO locks to the resonance of the other cavity. This is obtained simply by unlocking the loop, 
switching the error signal from $\epsilon_{S(1)}$ to $\epsilon_{S(2)}$ and re-lock. 
This operation is repeated periodically to verify the two numbers $n_{1,2}=\omega_0/(2 \pi \cdot {FSR_{1,2}})$. 
The auxiliary output of the VCO is connected to a microwave frequency counter (HP5343A) and the frequency counts are acquired by the PC via GPIB interface.
A gaussian fit on the distribution of the frequency counts provides the frequency mean value of the sample $f_{c}$ and its standard deviation $\sigma$. 
An example of dynamic resonance frequency estimation for the two cavities is presented in figure~\ref{gaussiane}.  

\subsection{Length-difference measurement}
Let's consider now the accuracy on the estimate of the length difference between two cavities having the same kind of mirrors. In this case, the phase 
correction terms in equation~\ref{EQ1} are equal and the mode number difference $n_D$:
\begin{equation*}
n_D=\frac{\omega_0}{2 \pi}\cdot \left(\frac{1}{FSR_2}-\frac{1}{FSR_1} \right)
\end{equation*}
should be an integer number. Figure~\ref {nD} shows the distribution of $n_D$ obtained from the measurements of figure~\ref{gaussiane}. 
The estimated mean value of the mode number difference is $ n_{D}=7427.4\pm1.6 $. The obtained uncertainty on $n_D$ sets the error on the length 
difference estimation to $\delta D=(\lambda /2) \cdot \delta n_D\sim 500 \,\,\rm {nm}$. 

\begin{figure}[h!!!]
\centering
\includegraphics[width=8.5cm]{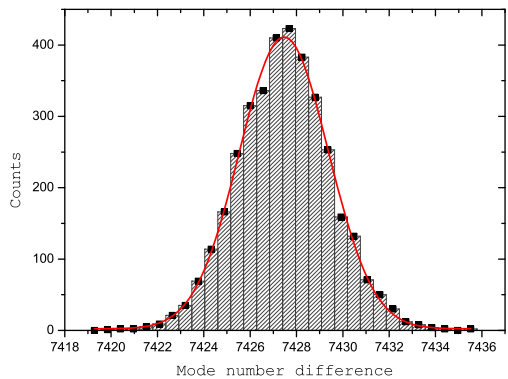}
\label{nD}
\caption{Distribution of the mode number difference $n_D$ obtained from the distribution of figure~\ref{gaussiane}. 
From the gaussian fit ($ \chi_R^{2}=0.997 $) we obtain $n_D=7427$ and $\sigma_{n_D}=1.6$}
\label{nD}
\end{figure}

\section{Conclusions}
We presented an experimental technique for the absolute length stabilization of the diagonal cavities formed by the opposite mirrors 
of large-frame He-Ne square ring lasers. 
As laser source for probing the cavities, a $ 10\,\,  \rm {mW} $ diode laser injection locked to an Iodine stabilized He-Ne laser has been developed. 
In typical operational conditions the system has a locking range of several $ \rm {GHz} $ and a power amplification of the order of 100. The large value of the locking range makes the system operation very robust against diode laser frequency drifts. The detection system makes use of a single EOM driven with three different modulation frequencies, allowing to lock the two cavities to the laser carrier and also to perform a detection of an integer multiple of the two free spectral ranges. 
We demonstrated a stable locking of the resonance frequencies of the two
cavities to the same reference laser, providing a length stabilization at the level of 
1 part in $10^{11}$ at 100 s, and the determination of the two FSRs to few parts in $10^7$. The error in the measurement of the cavity length difference with our set-up is of  $ 500\,\,  \rm {nm}$. 
We foresee the application of this technique to the GP2 prototype. For this ring laser the diagonal cavities will be contained inside a vacuum chamber and 
will have an expected finesse of $10^3$. In a temperature controlled laboratory, 
operating with a modulation frequency $\omega_B\sim 1\,\,\rm{GHz}$ (EOM bandwidth cut-off) we expect to be able to determine the mode number difference with 
an uncertainty less than one. In this case the error on the length difference between the two diagonals will be ultimately limited by the 
uncertainty on the laser wavelength.

\appendix
\section{}
Here we report some definitions recalled in the text. 
The reflection transfer function it is defined as:
\begin{equation}
\label{eq:Rtransf}
R(\omega)=\frac{\sqrt{r_{in}^{2}r_{out}^{2}}\left(e^{i\omega/FSR}-1\right)}{1-r_{in}^{2}r_{out}^{2}e^{i\omega/FSR}},
\end{equation}
where $r_{in}$, $r_{out}$ are the amplitude reflection coefficients for the input/output mirror, FSR is the cavity free spectral range and  $\omega$ the laser frequency.\\ 
The  amplitude and the frequencies of the nine components of the modulated laser field are listed in table~\ref{components}.
\begin{table}[h]
\label{components}
\centering
\begin{tabular}{|l|l|l|ll}
\cline{1-3}
k  & $A_k$	 			 & $\omega_k$ 		                                		&    \\ \cline{1-3}
1  &$ J_1(\alpha)  J_1(\beta)$    	 & $\omega_0-(\omega_B+\Delta\sin{\omega_C t})-\omega_A$ 		&    \\ \cline{1-3}
2  &$ -J_0(\alpha)  J_1(\beta)$           & $\omega_0-(\omega_B+\Delta\sin{\omega_C t})$ 			&    \\ \cline{1-3}
3  &$ -J_1(\alpha)  J_1(\beta)$  	 & $\omega_0-(\omega_B+\Delta\sin{\omega_C t})+\omega_A$ 		&    \\ \cline{1-3}
4  &$ -J_1(\alpha)  J_0(\beta)$  	 & $\omega_0-\omega_A$ 							&    \\ \cline{1-3}
5  &$ J_0(\alpha)  J_0(\beta)$  	 & $\omega_0$ 			        				&    \\ \cline{1-3}			
6  &$ J_1(\alpha)  J_0(\beta)$   	 & $\omega_0+\omega_A$ 	 						&    \\ \cline{1-3}			
7  &$ -J_1(\alpha)  J_1(\beta)$  	 & $\omega_0+(\omega_B+\Delta\sin{\omega_C t})-\omega_A$		&    \\ \cline{1-3}
8  &$ J_0(\alpha)  J_1(\beta)$ 	 	 & $\omega_0+(\omega_B+\Delta\sin{\omega_C t})$ 			&    \\ \cline{1-3}
9  &$ J_1(\alpha)  J_1(\beta)$   	 & $\omega_0+(\omega_B+\Delta\sin{\omega_C t})+\omega_A$ 		&    \\ \cline{1-3}
\end{tabular}
\caption{Definition of the nine amplitudes and frequencies of the electric field spectral components.}
\label{components}
\end{table}
\\

The components of the reflected intensity oscillating at $\omega_{A}$ comes
from the interference of the three groups of modes centered
respectively at $\omega_{0}(=\omega_{5})$, $\omega_{0}+\omega_{B}(=\omega_{8})$, $\omega_{0}-\omega_{B}(=\omega_{2})$:
\begin{eqnarray}
\label{omegaA}
I_{r}^{\omega_{A}} & \propto & 2E_{0}^{2}\{ A_{1}A_{2}\{ Re [S_{2}]\cos \omega At-Im [S_{2}]\sin \omega_{A}t\} +.\\\nonumber
 & + & A_{4}A_{5}\{ Re [S_{5}]\cos \omega_{A}t-Im [S_{5}]\sin \omega_{A}t\} \\\nonumber
 & + & .A_{7}A_{8}\{ Re [S_{8}]\cos \omega_{A}t-Im [S_{8}]\sin \omega_{A}t\} \} 
\end{eqnarray}
where $S_{i}$ is given by equation~\ref{eq:Si}. \\
For small values of $\frac{\Delta}{\omega_{B}}$, the component of detected intensity oscillating at $\omega_{C}$ is written as: 
\begin{eqnarray}
\label{omegaC}
I_{r}^{\omega_{C}} &\sim& E_{0}^{2}\Big[A_{1}^{2}\frac{d|R(\omega)|^{2}}{d \omega}\Big|_{(\omega=\omega_{1},\Delta=0)}+A_{2}^{2} \frac{d|R(\omega)|^{2}}{d\omega}\Big|_{(\omega=\omega_{2},\Delta=0)}\\\nonumber
&+&A_{3}^{2}\frac{d|R(\omega)|^{2}}{d\omega}\Big|_{(\omega=\omega_{3},\Delta=0)}+A_{7}^{2}\frac{d|R(\omega)|^{2}}{d\omega}\Big|_{(\omega=\omega_{7},\Delta=0)}\\\nonumber
&+&A_{8}^{2}\frac{d|R(\omega)|^{2}}{d\omega}\Big|_{(\omega=\omega_{8},\Delta=0)}+A_{9}^{2}\frac{d|R(\omega)|^{2}}{d\omega}\Big|_{(\omega=\omega_{9},\Delta=0)}
\Big]\Delta \sin {\omega_{C}t}.\nonumber
\end{eqnarray}

\section*{Acknowledgements}
The authors are grateful to F. Sorrentino and F. Stefani for the useful discussions.  We also acknowledge F. and M.
Francesconi, from the University of Pisa, for their help in the electronics development.
Special thanks go to F. Bosi and the ``Alte tecnologie" group of the INFN-Pisa for the fundamental mechanical support.


\section*{References}

\begin{thebibliography}{10}

\bibitem{Ullireview} Schreiber K U and Wells J-P R 2013 Large ring lasers for rotation sensing \textit{Rev. Sci. Instrum.} \textbf{84} (041101)1-26 doi:10.1063/1.4798216

\bibitem{stedreview} Stedman G E 1997 Ring-laser tests of fundamental physics and geophysics \textit{Rep. Prog. Phys.} \textbf{60} 615 doi:10.1088/0034-4885/60/6/001

\bibitem{tviolation} Stedman G E \textit{et al} 1995 T violation and microhertz resolution in a ring laser \textit{Opt. Lett.} \textbf{20} (3) 324--326

\bibitem{axions} Cooper L and Stedman G E 1995 Axion detection by ring lasers \textit{Phys. Lett. B} \textbf{357} 3

\bibitem{SCULLY} Scully M O, Zubairy M S 1981 Proposed optical test of metric gravitation theories \textit{Phys. Rev. A} \textbf{24} 4

\bibitem{stedman2003} Stedman G E, Schreiber K U, Bilger H R 2003 On the detectability of the Lense--Thirring field from rotating laboratory masses using 
ring laser gyroscope interferometers \textit{Class. Quant. Grav.}, \textbf{20} 2527-2540

\bibitem{GR_APPL} Chaboyer B and Henriksen R N 1988 Gravitational radiation observations with an orbital ring laser gyroscope \textit{Phys. Lett. A}, \textbf{132} 391-398

\bibitem{PRD_BOSI} Bosi F \textit{et al} 2011 Measuring gravitomagnetic effects by a multi-ring-laser gyroscope \textit{Phys. Rev. D} \textbf{84} (122002)1-23 doi:10.1103/PhysRevD.84.122002

\bibitem{controllodiG3} Schreiber K U \textit{et al} 2009 The large ring laser G for continuous Earth rotation monitoring \textit{Pure Appl. Geophys.} \textbf{166} 1485-1498 Birkhäuser Verlag ISSN 0033-4553 doi:10.1007/s00024-004-0490-4


\bibitem{controllodiG1} Schreiber K U, Gebauer, A, Wells J-P R 2013 Closed-loop locking of an optical frequency comb 
to a large ring laser \textit{Opt. Lett.} \textbf{38} 3574 - 3577 doi:10.1364/OL.38.003574

\bibitem{controllodiG2} Schreiber K U, Gebauer A, Wells J-P R 2012 Long-term frequency stabilization of a 16m2 ring laser gyroscope \textit{Opt. Lett.} \textbf{37} 1925-1927 doi:10.1364/OL.37.001925

\bibitem{G-Pisa} Belfi J \textit{et al} 2012 A 1.82 $m^{2}  $ ring laser gyroscope for nano-rotational motion sensing \textit{Appl. Phys. B} \textbf{106} 271-281 doi:10.1007/s00340-011-4721-y


\bibitem{GP2_1} Beverini N \textit{et al} 2014 Measuring general relativity effects in a terrestrial lab by means of laser gyroscopes \textit{Las. Phys.} \textbf{24} 074005

\bibitem{GP2_2} Belfi J \textit{et al} 2013 Absolute control of the scale factor in the GP2 laser gyroscope: toward a ground based detector of the Lense-Thirring effect\textit{ EFTF/IFC-21-25 July 2013, Prague} 795-798 doi:10.1109/EFTF-IFC.2013.6702268

\bibitem{Araya} Araya A \textit{et al} 1999 Absolute-length determination of a long-baseline Fabry-P\'erot cavity by means of resonating modulation sidebands \textit{Appl. Opt.} \textbf{38} 2848-2856 doi:10.1364/AO.38.002848

\bibitem{Aketagawa} Aketagawa M \textit{et al} 2010 Free Spectral Range
measurement of Fabry-P\'erot Cavity using Frequency Modulation \textit{IJPAP} \textbf{11} 851-856 doi:10.1007/s12541-010-0103-3

\bibitem{Aketagawa2} Aketagawa M \textit{et al} 2011 Measurement of a Free Spectral Range of a Fabry-P\'erot cavity using frequency modulation and null method under off-resonance conditions \textit{Meas. Sci. Technol.} \textbf{22} 025302 doi:10.1088/0957-0233/22/2/025302


\bibitem{Hood} Hood C J, Kimble H J and Ye J 2001 Characterization of high-finesse mirrors: Loss, phase shifts, and mode structure in an optical cavity \textit{Phys. Rev. A} \textbf{64} 033804, DOI: 10.1103/PhysRevA.64.033804

\bibitem{VANIO} Vainio M, Merimaa M and Nyholm K 2005 Optical amplifier for femtosecond frequency comb measurements near 633 nm \textit{Appl. Phys. B} \textbf{81} 1053-1057 doi:10.1007/s00340-005-1988-x

\bibitem{siegman} Siegman A E 1986 \textit{Lasers University Science Book} cap.29 ISBN 0-935702-11-5

\end{thebibliography}
\end{document}